\documentclass{PoS}

\newcommand{\vzero}{V~0332+53}

\title{The Be/X-ray binary system \vzero: A Short Review}

\ShortTitle{V0332+53: A Short Review}

\author{\speaker{M. D. Caballero-Garcia$^a$}, A. Camero-Arranz$^b$, M. \"{O}zbey Arabac{\i}$^{c}$, R. Hudec$^{ad}$, on behalf of a larger collaboration\\
\\
 \llap{$^a$} Czech Technical University in Prague, Faculty of Electrical Engineering \\
             Technick\'a 2, 166 27  Praha 6 (Prague), Czech Republic \\
        E-mail: \email{cabalma1@fel.cvut.cz}                \\
 \llap{$^b$} Institut de Ci\`{e}ncies de l'Espai, (IEEC-CSIC), Campus UAB, Fac. de Ci\`{e}ncies \\
             Torre C5 pa., 08193, Barcelona, Spain \\                                        
 \llap{$^c$} Department of Physics, Middle East Technical University \\
             Ankara, 06531, Turkey \\
 \llap{$^d$} Astronomical Institute, Academy of Sciences of the Czech Republic \\
             251~65~Ond\v{r}ejov, Czech Republic \\

}

\abstract{ Be/X-ray binary systems provide an excellent opportunity to study the physics around neutron stars through the study of the behaviour of 
matter around them. Intermediate and low-luminosity type outbursts are interesting because they provide relatively clean environments around neutron 
stars. In these conditions the physics of the magnetosphere around the neutron star can be better studied without being very disturbed by other phenomena
regarding the transfer of matter between the two components of the Be/X-ray binary system. A recent study 
presents the optical long-term evolution of the Be/X-ray binary {\vzero} plus the X-ray emission mainly during
the intermediate-luminosity outburst on 2008. In this paper we comment on the context of these observations and on the properties that can be derived through 
the analysis of them.  }

\FullConference{XI Multifrequency Behaviour of High Energy Cosmic Sources Workshop\\
		25-30 May 2015\\
		Palermo, Italy}

\begin{document}

\section{Introduction}

Accreting X-ray pulsars are binary systems composed of a donor star and an accreting neutron star (NS). In High Mass X-Ray
Binary (HMXB) systems the optical companion could be either a massive early-type supergiant (supergiant systems) or an
O,B main sequence or giant star (Be/X-ray binary; BeXB). Among the most remarkable signatures found in BeXBs are the detection of
IR excess and emission-line features in their optical spectra produced in a disc-like outflow around the Be star. Historically, their
outbursts have been divided into two classes. Type I (or normal) outbursts normally peak at or close to periastron passage of the
NS (L$_X\leq$ 10$^{37}$\,erg s$^{-1}$). Type II (or giant) outbursts reach luminosities of the order of the Eddington
luminosity (L$_X\sim$10$^{38}$\,erg s$^{-1}$; \cite{frank02}), with no preferred orbital phase. Also there are some ``intermediate''-luminosity 
states, referring to any X-ray outburst that does not comply with the aforementioned properties, thus have 
L$_X\sim$10$^{37}$-$10^{38}$\,erg s$^{-1}$ (\cite{caball15}).

\subsection{\vzero}

The recurrent hard X-ray transient {\vzero} was detected with the Vela 5B observatory in 1973 during its giant outburst, reaching a
peak intensity of ${\approx}1.6$\,Crab in the 3--12\,keV energy band \cite{terrell84}. The optical companion of the system, BQ~Cam \cite{argyle83} 
is an O8-9 type main sequence star at a distance of ${\approx}$7\,kpc \cite{negueruela99}. It has been widely observed both in optical 
and IR wavelengths since its identification. The optical spectrum is characterised by the highly variable
emissions of ${\rm H}_{\alpha}$, ${\rm H}_{\beta}$ and ${\rm H}_{\gamma}$ in addition to the He\,I lines. The photometric data show
an IR excess \cite{bernacca84,coe87,honeycutt85,unger92}.

\subsection{On the optical and X-ray emission}

Optical (spectroscopic and photometric) observations of BeXBs give us important information on the physics of the processes occurring in the 
outer envelope of the Be star. These stars are usually fast rotating and expel material into their surroundings in the form of a ``decretion'' 
disc surrounding them. The optical properties observed from these systems are dominated by the emission from this disc and its interaction
with the accretion disc surrounding the NS via mass-transfer. In the case of {\vzero} the latter is the less massive component of the 
binary system and spends most of its time in quiescence, when it is far away from the Be star. When it approaches 
the Be star during its orbit, then an X-ray outburst takes place (although there are BeXBs that show X-ray outbursts near the apastron passages 
of the NS). The X-ray emission results from the interaction of matter with the magnetosphere of the NS. This creates a very rich (and currently
not totally understood) phenomenology in the X-ray emission from these sources.

\section{History of the outbursts}

After its discovery in 1983 the system had passed a ten-year X-ray
quiescent phase when 4.4\,s pulsations were detected with Tenma and EXOSAT satellites \cite{stella85}. These X-ray activities, a series
of Type I outbursts, lasted about three months separated by the orbital period (34.25\,d) of the system. During these outbursts, rapid
random fluctuations in the X-ray emission in addition to the pulse-profile variations were also reported \cite{unger92}. The system
underwent another outburst, classified as Type II, in 1989 which led to the discovery of a cyclotron line scattering feature (CRSF) at 28.5\,keV
and two Quasi Periodic Oscillations (QPOs) centred at 0.051\,Hz and 0.22\,Hz \cite{makishima90,takeshima94,qu05}.

The brightening in optical/IR light-curves are
usually accompanied with the X-ray outburst phases as in the case of the giant 2004 outburst of the system. \cite{goranskij04} predicted this
outburst based on the optical brightening of the source in optical/IR band. During the outburst, three additional CRSFs at 27, 51 and 74\,keV were
detected in {\it Rossi X-ray Timing Explorer} ({\it RXTE}) observations \cite{coburn05} and confirmed by the subsequent {\it INTEGRAL} data
\cite{pottschmidt05}. \cite{tsygankov06} showed that the energy of these features is linearly anti-corrrelated with the luminosity of the source
indicating different X-ray states. The following X-ray activities of the system were in 2008, 2009 and 2010 with relatively weaker peak fluxes
\cite{krimm08,krimm09,nakajima10}. The system was in X-ray quiescence until 2015 June 18, when the Be companion probably reached its
maximum optical brightness after renewing activity in 2012 \cite{cameroarranz15}.

\begin{figure*}
\centering
\includegraphics[angle=0,width=0.8\linewidth]{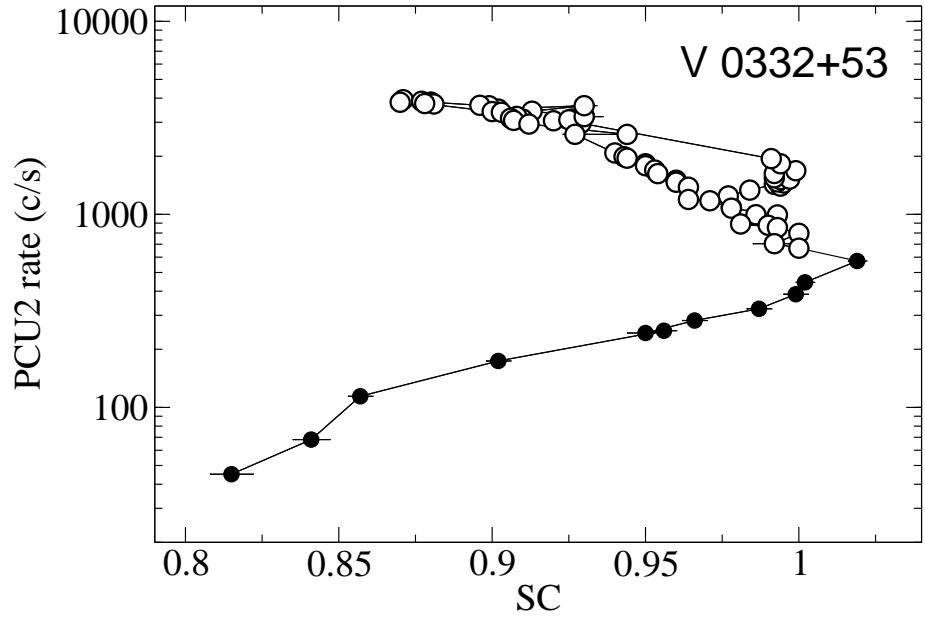}
\caption{Hardness (soft colour)-intensity diagram. The soft colour was defined as the ratio 7-10 keV / 4-7 keV. The count rate was obtained
      for the 4-30 keV energy band. Open circles designate points in the DB, while filled circles correspond to the HB. A
      logarithmic scale is used for the count rate. Figure from \cite{reig13}. }
\label{fig1}
\end{figure*}

A recent study of {\vzero} (\cite{reig13}) of giant outbursts at different luminosities has provided crucial insights into the most frequent
states of this source. These states are the low and high-luminosity ones. In Fig.~\ref{fig1} we can 
see the existence of two separate regimes (the diagonal branch - DB - and the horizontal branch - HB -). As from \cite{reig13} they are 
separated by a critical luminosity (${\rm L}_{\rm crit}{\approx}1-4{\times}10^{37}$\,erg\,s$^{-1}$, which corresponds to a {\it RXTE}/PCU2 rate of 
${\approx}800\,{\rm c}\,{\rm s}^{-1}$ in Fig.~\ref{fig1}). Both regimes
are clearly well-populated by X-ray observations whilst observations around the critical luminosity are scarce. This critical luminosity
coincides with the so-called intermediate-luminosity state.

\subsection{The intermediate-luminosity outburst in 2\,008}

\begin{figure*}
\centering
\includegraphics[bb=14 14 649 456,angle=0,width=0.8\linewidth]{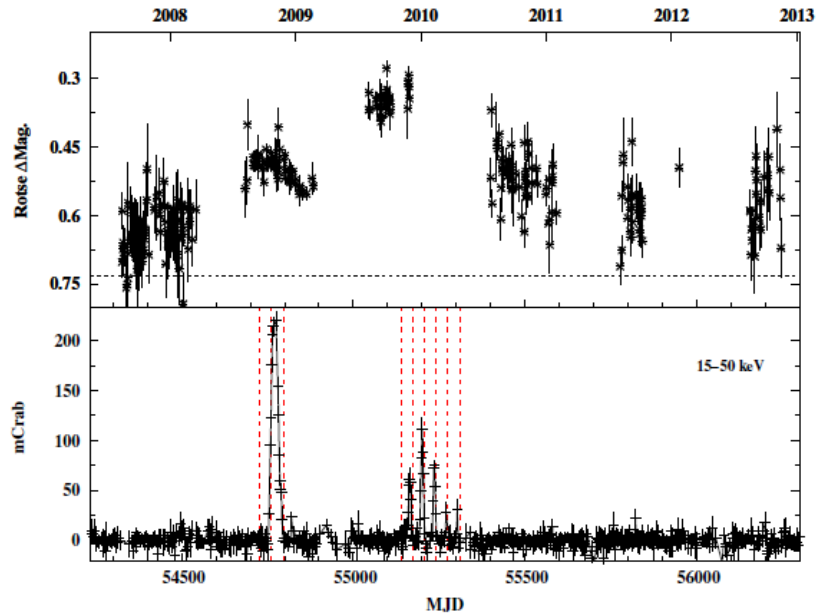}
\caption{Comparison of Swift/BAT light curve (15-50 keV) with a bin size equal to 2\,d with the ROTSEIII
(optical) magnitudes of {\vzero}, for the time interval MJD~54275-56300. The horizontal back-dashed
line in the ROTSE panel denotes quiescent differential magnitude of {\vzero}
whereas the vertical red-dashed lines in the X-ray panel represent the times of the periastron
passages of the NS. Figure adapted from \cite{caball15}.  }
\label{fig2}
\end{figure*}

A multi-wavelength study of {\vzero} mostly during the recent events initiated in 2008 has been recently performed (\cite{caball15}). For 
this purpose we used archived {\it Swift}-XRT and {\it RXTE} pointed observations carried out in 2008, as well as one {\it Suzaku} observation from 2010
and survey data from different space-borne telescopes and covering intermediate and low-luminosity events. In addition, we used optical/IR data from
our dedicated campaign involving several ground-based astronomical observatories (Fig.~\ref{fig2} shows a zoom of the optical and
X-ray light curve covering such time). This provides us a unique opportunity to undertake studies
of X-ray outbursts of intermediate-low X-ray luminosities. The former is a regime very scarcely explored in {\vzero}, one of the BeXBs with strongest
magnetic field already known.

\section{Discussion and conclusions}

Here we present a hint of the optical long-term evolution of the BeXB {\vzero} plus the 
X-ray emission mainly during the intermediate-luminosity 
outburst on 2\,008. Most of the characteristics found are alike those found in giant outbursts. Nevertheless, some particular points 
do not coincide which the aforementioned outbursts, e. g.: 

\begin{itemize}
\item{The presence of QPOs at a frequency close to the pulse of the NS during the lowest luminosities.}
\item{A positive correlation of the energy of the CRSF line with luminosity.}
\end{itemize}

We refer to \cite{caball15} for a report on the details and discussion of these (and other) results obtained.

\acknowledgments
MCG is supported by the European social fund within the framework of realizing the project ``Support of inter-sectoral mobility and
quality enhancement of research teams at Czech Technical University in Prague'', CZ.1.07/2.3.00/30.0034.

\end{document}